\documentclass[10pt]{llncs}

\usepackage{placeins} 
\usepackage{underscore} 
\usepackage[font=itshape]{quoting} 
\usepackage{graphicx} 
\usepackage{epstopdf} 
\usepackage{textcomp} 
\usepackage{amsmath} 
\usepackage{pbox} 
\usepackage{hyperref} 
\usepackage{cite} 
\usepackage{url} 
\usepackage[misc]{ifsym} 

\newcommand*\justify{
  \fontdimen2\font=0.4em
  \fontdimen3\font=0.2em
  \fontdimen4\font=0.1em
  \fontdimen7\font=0.1em
  \hyphenchar\font=`\-
}

\usepackage{array}
\usepackage{booktabs}
\usepackage{multirow}

\usepackage{amssymb}
\usepackage{pifont}

\usepackage{tabularx}
\usepackage{threeparttable, tablefootnote}
\usepackage{colortbl} 

\usepackage{msc} 
\usepackage{xcolor}
\newcommand{\quotes}[1]{``#1''} 
\begin{document}

\title{On the Feasibility of Fine-Grained TLS Security Configurations in Web Browsers Based on the Requested Domain Name}

\author{Eman Salem Alashwali\inst{1,}\inst{2}\textsuperscript{ (\Letter)} \and Kasper Rasmussen\inst{1}}  
\institute{University of Oxford, Oxford, United Kingdom \\
\email{\{eman.alashwali,kasper.rasmussen\}@cs.ox.ac.uk} \\
\and King Abdulaziz University (KAU), Jeddah, Saudi Arabia \\
\email{ealashwali@kau.edu.sa}
}
\maketitle
\setcounter{footnote}{0}

\begin{abstract} \label{abstract}
Most modern web browsers today sacrifice optimal TLS security for backward compatibility. They apply coarse-grained TLS configurations that support (by default) legacy versions of the protocol that have known design weaknesses, and weak ciphersuites that provide fewer security guarantees (e.g. non Forward Secrecy), and silently fall back to them if the server selects to. This introduces various risks including downgrade attacks such as the POODLE attack \cite{moller14} that exploits the browsers silent fallback mechanism to downgrade the protocol version in order to exploit the legacy version flaws. To achieve a better balance between security and backward compatibility, we propose a mechanism for fine-grained TLS configurations in web browsers based on the sensitivity of the domain name in the HTTPS request using a whitelisting technique. That is, the browser enforces optimal TLS configurations for connections going to sensitive domains while enforcing default configurations for the rest of the connections. We demonstrate the feasibility of our proposal by implementing a proof-of-concept as a Firefox browser extension. We envision this mechanism as a built-in security feature in web browsers, e.g. a button similar to the \quotes{Bookmark} button in Firefox browsers and as a standardised HTTP header, to augment browsers security.
\end{abstract}


\section{Introduction} \label{introduction}
The Transport Layer Security (TLS) protocol \cite{tls12}\cite{tls13rev24} is one of the most important and widely used protocols to date. It is used to secure internet communications for billions of people everyday. TLS provides a secure communication channel between two communicating parties. At the beginning of each new TLS session, the client and server must agree on a single common TLS version and ciphersuite to be used in that session. These are extremely important parameters as they define the security guarantees that the protocol can provide in a particular session. TLS supports various protocol versions and ciphersuites. Each ciphersuite is a string that defines the cryptographic algorithms that will be used in a particular session. Generally, these algorithms include\footnote{TLS~1.2 and TLS~1.3 ciphersuite strings have different format and define different set of algorithms. See \cite{tls12} and \cite{tls13rev24} for more details.}: the Key-Exchange, Digital Signature, Symmetric Encryption, Authenticated Encryption (AE), and Hash. Clients\footnote{In our paper, TLS clients are represented by web browsers. We will use the terms client, web browser, or browser interchangeably.} and servers tend to support legacy versions, e.g. TLS~1.1 and TLS~1.0, and weak, less secure, or unrecommended ciphersuites, e.g. non-Forward Secrecy (non-FS), non-Authenticated Encryption (non-AE) \footnote{In our context, non-AE refers to ciphersuites that do not provide confidentiality, integrity, and authenticity simultaneously. For example the CBC MAC-then-encrypt ciphersuites which are susceptible to padding oracle attacks \cite{cloudflare16}\cite{vaudenay02}.}, or weak Hash algorithms, mainly to provide backward compatibility with legacy servers. For example, a recent analysis of IPv4 internet scan dataset shows that embedded web servers in networked devices tend to use legacy TLS versions compared to top domain web servers \cite{samarasinghe17}. To accommodate such servers, web browsers tend to support legacy versions and weak ciphersuites to be able to connect to such legacy servers. The same goes for updated servers. They support legacy versions and weak ciphersuites so as not to lose connections from legacy clients. \par

\subsection{Motivation}
In the near future, the coming version of TLS, TLS~1.3, which is currently work in progress \cite{tls13rev24}, will become a standard. Ideally, mainstream browsers\footnote{Throughout the paper, mainstream browsers refer to the following tested versions: Chrome version 63.0.3239.108, Firefox 57.0.2, Internet Explorer 11.125.16299.0, Edge 41.16299.15.0, and Opera 49.0.2725.64.} will deploy it and offer it as the default version. TLS~1.3 provides significant improvements in security and performance over its predecessors. However, experience has shown that ordinary web servers may take years till they get upgraded to support the latest TLS version. This is especially true in embedded web servers as we mentioned earlier. For this reason, web browsers tend to maintain support for legacy TLS versions and weak ciphersuites and silently (without warning or indicator) fall back to them if the server they are trying to connect to does not support the latest version or the strongest ciphersuite. This is the case in all mainstream web browsers today. It is not inconceivable that this will remain the case after the deployment of the coming version, TLS~1.3, despite numerous known weaknesses including design flaws in the current version, TLS~1.2. For example, legacy versions up to TLS~1.2 do not authenticate the server\textquotesingle s selected version and ciphersuite at early stage of the handshake. This flaw allows various attacks that result in breaking the protocol\textquotesingle s main security guarantees as shown in \cite{adrian15}\cite{beurdouche15}, for example.\par

There are several TLS attacks that exploit the support for legacy versions or weak ciphersuites by one or both of the communicating parties during the TLS handshake. This family of attacks is known as downgrade attack, where an active network attacker forces the communicating parties to operate in a mode that is weaker than they would prefer and support, in order for him to perform attacks that would not have been possible in the strong mode. For example, \cite{adrian15}\cite{aviram16}\cite{beurdouche15}\cite{beurdouche14}\cite{moller14}, among others.

The tension between security and backward compatibility in cryptographic protocols is historical. Backward compatibility seems inevitable in internet protocols such as TLS due to the global and heterogeneous nature of the connected devices over the internet, in addition to the heavy reliance on the internet services in people\textquotesingle s daily lives. From the client\textquotesingle s perspective (which is our focus in this paper), if a browser is configured with optimal TLS configurations, i.e. it only negotiates the latest version of TLS, e.g. TLS~1.3, and a handful of the strongest ciphersuites that satisfy \textit{both} FS and AE properties, this can be the strongest client but can render many ordinary websites on legacy servers unreachable due to compatibility issues. Obviously, this will lead to a difficult user experience. On the other hand, if the client supports legacy versions, e.g. TLS~1.0 and weak or non preferred ciphersuites, e.g. non-FS or non-AE as is the case in all mainstream web browsers today, the browser silently falls back to one of the legacy versions or weak ciphersuites to connect to those legacy servers that do not support the latest version or the strongest ciphersuites. \par 

To maintain a balance between the two extremes, the browser needs to distinguish between various contexts and apply fine-grained TLS configurations. That is, the browser enforces optimal configurations for connections going to sensitive domains, and default ones for the rest of the connections. \par 

To this end, we try to answer the following question: \textit{How can we guide the browser into making an informed decision on whether to enforce optimal or default TLS configurations?} \par

\subsection{Contribution} \label{our contribution}
Our contribution is twofold: First, we propose a light-weight mechanism for fine-grained TLS security configurations in web browsers. Our mechanism allows browsers to enforce optimal TLS security configurations for connections going to sensitive domains while maintaining default configurations for the rest of the connections. It represents a middle-ground between optimal TLS configurations that might render many ordinary websites unreachable and default configurations that might be abused by attackers to perform downgrade attacks. Our mechanism can detect and prevent a class of dangerous downgrade attacks and server misconfigurations. Furthermore, it does not require a new Public Key Infrastructure (PKI) nor third parties such as Certificate Authorities (CAs). Second, we examine the feasibility of our mechanism by implementing a proof-of-concept as a Firefox browser extension. In addition, we present the extension architecture. \par

\subsection{Organisation} \label{organisation}
The rest of the paper is organised as follows: In section \ref{background} we provide a brief background. In section \ref{related work} we summarise some related work. In section \ref{threat model} we describe our system and threat models, and goals. In section \ref{proposal} we present our proposed mechanism. In section \ref{other methods} we briefly describe some other server-based TLS configurations advertisement methods. In section \ref{limitations} we list some limitations. Finally, in section \ref{conclusion} we conclude.

\section{Background} \label{background}

\subsection{TLS Version and Ciphersuite Negotiation} \label{tls negotiation}
\begin{figure}[!tp]
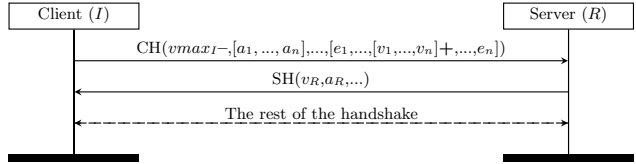
 
\centering
\resizebox{0.7\textwidth}{!}{
\begin{msc}[small values, /msc/level height=0.6cm, /msc/label distance=0.5ex, /msc/first level height=0.6cm, /msc/last level height=0.6cm]{} 
\setmsckeyword{} 
\drawframe{no} 

\setlength{\instwidth}{2.5\mscunit} 
\setlength{\instdist}{7\mscunit} 

\declinst{I}{}{Client ($I$)}
\declinst{R}{}{Server ($R$)}

\mess{CH($vmax_I$\textbf{--},[$a_1,...,a_n$],...,[$e_1$,...,[$v_1$,...,$v_n$]\textbf{+},...,$e_n$])} {I}{R}
\nextlevel

\mess{SH($v_R$,$a_R$,...)} {R}{I}
\nextlevel

\mess*{The rest of the handshake}{I}{R}
\mess*{}{R}{I}

\end{msc}
} 

\caption{Simplified message sequence diagram illustrating the version and ciphersuite negotiation in the TLS \texttt{Hello} messages. Parameters followed by \quotes{--} are deprecated in TLS~1.3 while those followed by \quotes{+} are newly introduced in TLS~1.3. The unmarked parameters are mutual to both versions.}
\label{fig:TLS-CH} 
\end{figure}

We now briefly describe the version and ciphersuite negotiation in both TLS~1.2 \cite{tls12} and TLS~1.3 (draft-24) \cite{tls13rev24}. We base our description on the current version, TLS~1.2, and if there is any difference in TLS~1.3 we mention it explicitly. As depicted in Figure \ref{fig:TLS-CH}, at the beginning of a new TLS handshake the client (Initiator \textit{I}) must send a \texttt{ClientHello} (\texttt{CH}) message to initiate a connection with the server (Responder \textit{R}). The \texttt{ClientHello} contains several parameters including: First, the client\textquotesingle s TLS supported versions which is sent as a single value that represents the maximum supported version ($vmax_I$) while in TLS~1.3, it is sent as a list of supported versions ([$v_1,...,v_n$]) in the \quotes{supported_versions} extension. The $vmax_I$ is still included in TLS~1.3 \texttt{ClientHello} for backward compatibility and its value is set to TLS~1.2. Second, a list of supported ciphersuites ([$a_1,...,a_n$]). Third, a list of the client\textquotesingle s supported extensions ([$e_1$,...,$e_n$]) is sent at the end of the message. In TLS~1.3 the extensions must at least include the \quotes{supported_versions} [$v_1,...,v_n$], while in TLS~1.2, the extensions are optional.\par

Upon receiving a \texttt{ClientHello}, the server decides which version and ciphersuite will be used in the session and responds with a \texttt{ServerHello} (\texttt{SH}). The \texttt{ServerHello} contains several parameters including: First, the server\textquotesingle s selected TLS version ($v_R$) based on the client\textquotesingle s supported versions. Second, the selected ciphersuite ($a_R$) based on the client\textquotesingle s proposed list. If the server does not support any of the client\textquotesingle s proposed versions or ciphersuites, it responds with a handshake failure alert. However, if the server selected a version lower than the client\textquotesingle s maximum version, all mainstream web browsers today fall back silently to a lower version (up to TLS~1.0). 


\subsection{TLS Downgrade Attack} \label{defense}
In recent years, several downgrade attacks have been shown practical. For example, the version downgrade in the Padding Oracle On Downgraded Legacy Encryption (POODLE) attack \cite{moller14}, the \quotes{Version rollback by \texttt{ClientHello} fragmentation} \cite{beurdouche14}, and the ciphersuite downgrade (from RSA to non-RSA) in a variant of the DROWN attack \cite{aviram16}. \par
 
\begin{figure}[!tp]
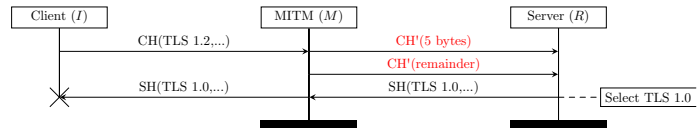
 
\centering
\resizebox{0.8\textwidth}{!}{
\setmsckeyword{} 
\drawframe{no} 
\begin{msc}[normal values, /msc/level height=0.6cm, /msc/label distance=0.5ex , /msc/first level height=0.6cm, /msc/last level height=0.6cm, /msc/top head dist=0, /msc/bottom foot dist=0]{}
\setlength{\instwidth}{2.5\mscunit} 
\setlength{\instdist}{4\mscunit} 
\declinst{I}{}{Client ($I$)}
\declinst{M}{}{MITM ($M$)}
\declinst{R}{}{Server ($R$)}
\mess {CH(TLS~1.2,...)}{I}{M}
\mess {\textcolor{red}{CH\textquotesingle{}(5 bytes)}}{M}{R}
\nextlevel
\mess {\textcolor{red}{CH\textquotesingle{}(remainder)}}{M}{R}
\nextlevel
\msccomment[r]{Select TLS~1.0}{R}  
\mess {SH(TLS~1.0,...)}{R}{M}
\mess {SH(TLS~1.0,...)}{M}{I}
\stop{I}
\end{msc}
} 
\caption[caption]{Illustration of a version downgrade attack attempt based on the \quotes{Version rollback by ClientHello fragmentation} attack scenario \cite{beurdouche14} when the client does not support legacy TLS versions.} 
\label{fig:version rollback protection}
\end{figure}

The aforementioned attacks share a pattern: First, the client supports either a legacy version (SSL~3.0 or TLS~1.0) or non preferred or weak ciphersuite (RSA) and silently falls back to them. Second, the attacks circumvent the handshake transcript authentication mechanism (in the \texttt{Finished} MACs) that is placed to detect any modifications in the protocol messages (including the version or ciphersuite). Clearly, these attacks could have been prevented if the client does not support legacy versions or weak ciphersuites. In these cases, the client will refuse to proceed the handshake with a version or ciphersuite that it does not support, as illustrated in Figure \ref{fig:version rollback protection}. One might argue that this is what all browsers should do: disable all legacy versions and unrecommended ciphersuites that provide fewer security guarantees (e.g. non-FS) and never accept them from any server. Experience shows that disabling legacy versions and exclusively offering the strongest ciphersuites is a complex decision for browser vendors. Browser vendors scarifies optimal TLS security configurations to some degree to provide backward compatibility for their users who might need to connect to legacy servers as we mentioned earlier. However, in this paper, we do not argue for or against. Rather, we explore the solutions space to augment browsers security while maintaining usability and backward compatibility. Our mechanism augments browsers security for connections to sensitive domains that are capable of providing optimal TLS configurations (but also support legacy configurations for legacy clients) by providing the browser with prior knowledge about these sensitive domains. This is an improvement over the \quotes{one-size-fits-all} TLS security policy in all mainstream web browsers which renders some downgrade attacks undetected.  

\section{Related Work} \label{related work}

\subsection{Browsers Security Enhancement Mechanisms}
In \cite{jackson08} Jackson and Barth introduce \quotes{ForceHTTPS}, a mechanism for enforcing strict HTTPS policy. Websites opt-in to this policy either from the server side by advertising a special HTTP response header, or from the client side by the user. The strict policy converts any plain HTTP URL to HTTPS and performs stricter TLS certificate validation than the default one. The mechanism blocks any opted-in website that violates the policy. Our work is similar in spirit to \cite{jackson08}, and can be viewed as an extension to \cite{jackson08}, at a finer level. Our mechanism is for enforcing optimal TLS \textit{configurations} (TLS version and ciphersuites) using a different technique (whitelisting). Unlike a decade ago, nowadays enforcing TLS is not sufficient. We aim for methods to enforce optimal TLS. Experience has shown the practicality of TLS version and ciphersuite downgrade attacks that circumvent design-level downgrade protection mechanisms (e.g. by exploiting implementation bugs). These attacks abuse the support of legacy versions or ciphersuites by one or both parties, which result in breaking TLS main security guarantees as in \cite{aviram16}\cite{beurdouche14}\cite{moller14}. \par

In \cite{policert14} Szalachowski et al. propose PoliCert which introduces the idea of policy certificate which includes optional parameters to specify the minimum TLS security level that the client should enforce and the error handling mechanism. Our idea of enforcing fine-grained TLS configurations complements and provides a new perspective for a relevant concept. However, there are subtle differences between the two work. For example, unlike \cite{policert14}, our proposal does not require a new PKI nor CAs. It is a light-weight mechanism that can be adopted by browser vendors as an additional layer of security. \par
Overall, we are not aware of an existing light-weight mechanism or browser extension that proposes or implements our concept in this paper. 

\subsection{Warning Messages in Web Browsers}
Previous work shows that users tend to ignore passive security indicators such as the padlock and the Extended Validation (EV)\footnote{Extended Validation is a passive browser indicator that appears in the address bar only for websites which have strongly verified identity.} indicators \cite{jackson07}\cite{schechter07}.\par

On the other hand, it has become clear that active warnings that interrupt the user\textquotesingle s task and ask for the user\textquotesingle s action are more effective than passive indicators \cite{akhawe13}\cite{egelman08}\cite{schechter07}\cite{wu06}. However, users adherence to a warning vary with context such as site reputation \cite{reeder18}. The study suggests considering contextual factors to improve warning messages \cite{reeder18}. Our mechanism and previous work such as ForceHTTPS \cite{jackson08} and PoliCert \cite{policert14} suggest giving server administrators and domain owners means to advertise strict TLS security policies and error handling which can help protect users from making bad security decisions. \par
    
Warning messages should be avoided in benign situations to avoid the \quotes{habituation} effect that results from seeing the warning too often such that users underestimate the risk behind it \cite{sunshine09}. \par

These studies give useful insights that will be considered in the usability aspect of our mechanism. 

\section{Our System and Threat Models, and Goals} \label{threat model}
Our system model consists of TLS client and server trying to establish a TLS connection. TLS proxies\footnote{A proxy (also known as middlebox) is an entity that can be placed between a client and server for various purposes such as interception or packet inspection. It splits the TLS connection between the client and server so that the client and server are in fact connecting to the proxy and not directly to each other.} (middleboxes) are out of our system\textquotesingle s model scope. Both parties support multiple TLS versions and ciphersuites with various security levels. The TLS client is represented by mainstream web browsers. It supports and prefers the latest TLS version and the strongest ciphersuite. However, for backward compatibility, it also supports legacy versions and weak ciphersuites, and silently falls back to them if the server selected a legacy version or weak ciphersuite. The client implements the \quotes{downgrade dance} mechanism that makes the browser fall back to a lower version and retry the handshake if the initial handshake failed for any reason as is the case in the POODLE attack \cite{moller14}. Ideally, updated web browsers today do not support completely broken cryptographic algorithms (ciphersuites). However, they do support ciphersuites that provide fewer security guarantees such as those with non-FS key-exchange or non-AE. The assumption that the browser supports weak, unrecommended, or plausibly broken ciphersuites is realistic and can happen in the future. Classical cryptographic algorithms do not last forever. Algorithm design flaws can be found, and advances in computation powers enable solving hard problems such as prime factorisation. For example, several algorithms were supported through years of speculations about their insecurity until they got officially deprecated such as the RC4 algorithm \cite{rc415}.  \par 

In terms of threat model, our model assumes that the client and server are honest peers. The adversary has full control over the communication channel and can drop, modify, inject, or redirect messages in the channel. \par 

In terms of system goals, our proposal tries to mitigate the risks that can result from the browsers silent fallback to a legacy TLS version or weak ciphersuite which puts the client at the risk of: 

\begin{enumerate}
\item Falling victim to downgrade attacks by a man-in-the-middle that exploits the client support for legacy versions or ciphersuites. For example, the case of version downgrade in the POODLE attack \cite{moller14} which allows the attacker to exploit flaws in the legacy version to decrypt secret data.

\item Connect to misconfigured servers for important services such as ebanking and egovernment websites, or important services that are not necessarily maintained by large service providers or experts who are up-to-date with advances in security. For example, an organisation\textquotesingle s web mail server. 
\end{enumerate}
     
\section{Our Proposal} \label{proposal}

\subsection{Overview} \label{overview}
\begin{figure*}[!tp]
\centering
\includegraphics[width=\textwidth]{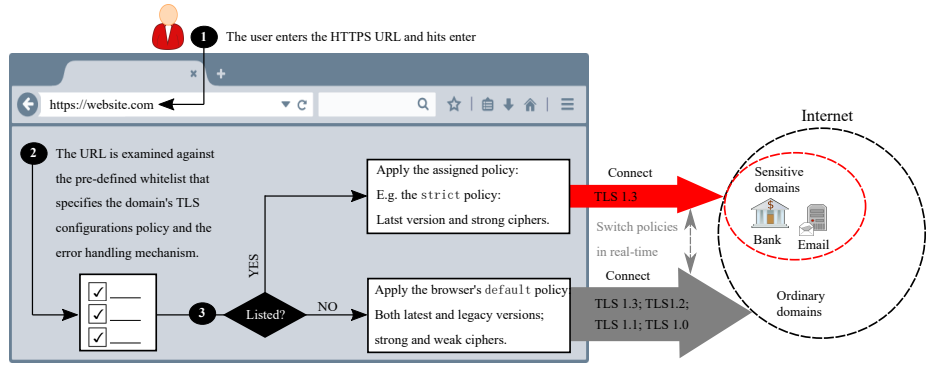}
\caption{Illustration for our proposed fine-grained TLS configurations mechanism applied on the TLS versions (omitting the ciphersuites for simplicity).}
\label{fig:sysoverview}
\end{figure*}

Our proposal tries to tackle the challenge of providing a high level of security for connections to sensitive websites while maintaining backward compatibility with ordinary websites in TLS implementations in web browsers. To this end, we propose a mechanism for fine-grained TLS configurations. That is, optimal TLS configurations are enforced for connections to sensitive domains, while default configurations are enforced for the rest of the connections. To do this, the browser needs guidance to distinguish between different contexts. We achieve this by providing the browser with prior knowledge through a pre-defined list of sensitive domain names, e.g. ebanking, egovernment, ebusiness portals, etc. that guides the browser into whether to enforce the optimal TLS configurations or the default ones. See Figure \ref{fig:sysoverview} for an overview of our proposed mechanism. \par 

To realise the idea, we implemented a proof-of-concept as a Firefox browser extension. For the extension\textquotesingle s implementation, we built a hybrid Firefox extension using WebExtensions API \cite{webextensions} and Add-on SDK \cite{sdk17}. We run the extension in Firefox Developer edition version 56.0b3 (32-bit). The overall concept seems straight-forward but implementing it required non-trivial effort to overcome WebExtensions API limitations to perform low-level functions such as the configurations re-writing and error messages customisation.\par

\subsection{Subscribing to the Mechanism}
Websites (domains) can subscribe to our mechanism by two methods. The first method is a client-side method which targets two groups of users: First, security-conscious users who want to protect their sensitive connections, e.g. connections to email and ebanking websites, from the threats described in the system goals in section \ref{threat model}. Second, Information Technology (IT) administrators in organisations such as banks, universities, etc. who can whitelist some sensitive domains in their employees PCs and laptops so that any connection from their devices to these domains is initiated using optimal TLS configurations. Similar to any client-side security feature, the mechanism requires some level of awareness about its benefits to incentivize users to use it. \par 

The second subscription method is a server-side method through an HTTP response header. The server administrator needs to configure the server to send the mechanism\textquotesingle s specific header. Upon receiving, our extension adds the domain name that has sent the header to the whitelist. This method does not require user intervention but assumes an authentic first connection that contains the HTTP header, also known as Trust On First Use (TOFU). This is a well understood prerequisite for all header-based policies such as HSTS \cite{hsts17}, Content Security Policy (CSP) \cite{csp10} and Public-Key Pinning (PKP) \cite{pinning15}. Therefore, the user must make the first connection to these website that advertise the mechanism\textquotesingle s header using a trusted network. In addition, unlike the client-side method where the policy can be enforced before the first HTTPS request is sent, in the server-side subscription method, the policy can only be enforced after a first request is sent (to get the server response header first).    

\subsection{Architecture} \label{architecture}
Our Firefox extension\textquotesingle s components can be described as follows:

\subsubsection{Pre-Defined TLS Configuration Policies.} \label{policies}
The policies govern the TLS configurations that will be enforced before an HTTPS request is sent. The TLS configurations space for our policies is the set of versions and ciphersuites that exist in Firefox Developer edition version 56.0b3. That is, in terms of versions: TLS~1.3, TLS~1.2, TLS~1.1, and TLS~1.0, and in terms of ciphersuites: 15 ciphersuites that provide different levels of security guarantees which include FS, non-FS, AE, and non-AE ciphersuites. There is a consensus among the security community that FS and AE ciphersuites provide stronger security guarantees than non-FS or non-AE. Therefore, for our set of ciphersuites in Firefox we can define strong cipehrsuites as those that provides \textit{both} FS and AE, while weak ciphersuites as those that do not support FS or AE or both, as they provide fewer security guarantees. Similarly, if there are known or plausibly broken primitives in any cipehrsuite (this can happen in the future), it can not be added to the set of strong ciphersuites.\par 

For our proof-of-concept, we define two TLS configuration policies and hard code them in the extension: \texttt{strict} which represents the optimal TLS configurations, and \texttt{default} which represents the default configurations. The \texttt{strict} policy is the strictest class which supports only TLS~1.3\footnote{We hypothetically and proactively assume TLS~1.3 is the highest version in our TLS policy levels\textquotesingle{} definitions. This will be the case when TLS~1.3 becomes a standard soon. However, in practice (and in our proof-of-concept) the highest possible version is still TLS~1.2.}, and only the TLS~1.3 ciphersuites that provide \textit{both} AE and FS. Second, the \texttt{default} policy (default configurations) is equivalent to the Firefox Developer edition version 56.0b3 default configurations which support TLS~1.3, TLS~1.2, TLS~1.1 and TLS~1.0, and 15 different ciphersuites including FS, non-FS, AE, and non-AE ciphersuites. See Table \ref{table:levels} for our proof-of-concept TLS policy levels.\par
 
The number of policy levels is a design decision. We could have defined more levels if desired. For example, we could have defined a level for each version of TLS. The more levels, the more granularity is achieved. However, more granularity means more warning messages since the essence of our mechanism is to either warn the user before falling back to a lower policy or block the user from proceeding to the website (depending on the error handling mechanism that is assigned to that domain as we will elaborate later), unlike the browser\textquotesingle s default behaviour that silently falls back to a lower version or weaker ciphersuite without warning which can render some downgrade attacks or misconfigured important servers undetected. We try to maintain a balance between security and usability and decided to define two levels policy where there is a significant shift in the provided security guarantees, and present only one warning message to the user in case of policy violation.

\begin{table}[!t]
\centering
\noindent
\caption{Our extension\textquotesingle s built-in policies.}
\label{table:levels}
\begin{tabular*}{0.9\textwidth}{lllllll}
\\ \toprule
Policy Level   &&& TLS Version  				&&& Ciphersuites \\  \midrule
Strict	&&& TLS~1.3	 							&&& \textit{Both} FS and AE\\ 
Default &&& TLS~1.3; TLS~1.2; TLS~1.1; TLS~1.0	&&& FS; AE; non-FS; non-AE\\
\bottomrule \\
\end{tabular*}
\end{table}  

\subsubsection{Pre-Defined Domain Names List.} \label{domain list}
The domain names list combined with the TLS policy that is assigned to each domain name is used to provide the browser with the prior knowledge that guides it into which TLS policy should be enforced for each examined HTTPS request. In our proof-of-concept, the domain names list takes the domain names from two sources: either manually as an input from the user, or automatically by extracting the domain name from the URL that sent the mechanism\textquotesingle s specified HTTP header. The domains are entered in the form of \quotes{example.com}. We do not allow duplicate domain names. Therefore, a domain\textquotesingle s TLS policy can not be over-written unless after removing the existing record. By default, the domain names are added to the strictest TLS policy that enforces optimal TLS configurations which is the \texttt{strict} policy. However, if the connection using the optimal TLS configurations could not be established due to lack of server support for the requested configurations, the user is presented with a warning message. Depending on the error handling mechanism (will be explained next) that is assigned to that particular domain, the user will be either blocked from proceeding to the website, or warned and allowed to relax the domain\textquotesingle s TLS policy to a lower one (from \texttt{strict} to \texttt{default} in our case). 

\subsubsection{Pre-Defined Error Handling Mechanism.}
In general, there are three main strategies for error handling in web browsers: blocking, active warning, and passive warning. Blocking is a conservative approach that blocks the user from proceeding to the website. It should be used when the attack is certain. This approach is adopted by the \quotes{ForceHTTPS} mechanism which considers violating the strict TLS policy by an opted-in website as an attack \cite{jackson08}. On the other hand, the active warning strategy is less conservative. It temporarily blocks the user to warn him, but it allows him to click-through (bypass) the error through one or multiple clicks. This strategy is used in the self-signed certificate warning in most browsers today. Finally, the passive warning strategy shows an indicator which can be negative or positive indicator without interrupting the user\textquotesingle s task, e.g. the padlock icon. As stated earlier, previous studies suggest that active warnings are more effective than passive ones, but need to be used with caution not to cause the \quotes{habituation} effect. \par

In our browser extension, the error handling mechanism specifies the type of the error message that will be presented to the user in case of TLS policy violation. Each whitelisted domain has a TLS policy and an error handling mechanism assigned to it. We define two error handling mechanisms: \texttt{blocking} and \texttt{active warning}. The error mechanism depends on the subscription method (client-side or server-side) which implies the level of confidence on the server\textquotesingle s ability to meet optimal TLS configurations. 

\begin{itemize}
\item If the domain subscription is client-side through a user, the user has the choice to assign either \texttt{blocking} or \texttt{active warning} error handling to the domain. By default, the error handling mechanism for client-side subscription is set to \texttt{active warning}. However, if the user (e.g. IT administrator) has high level of confidence that the added domain should be able to meet the \texttt{strict} TLS policy (e.g. bank or enterprise server), he can select the \texttt{blocking} error handling mechanism to block the user from proceeding to the website if the TLS server response violated the policy. 

\item If the domain subscription is server-side through an HTTP response header, the mechanism automatically assign the \texttt{blocking} error handling mechanism for the advertising domain. Servers that advertise the mechanism\textquotesingle s header must first ensure that they are capable of meeting the \texttt{strict} TLS policy requirements. They must be aware that their users will be blocked from reaching the server if the \texttt{strict} TLS configurations policy has been violated. This is a conservative approach towards highly secure connections and reduced decision making effort on users, in the same direction of HSTS policy \cite{hsts17}. We adopt it when the confidence of the server\textquotesingle s TLS capabilities is high (i.e., when the knowledge comes from the server side) to avoid denial of service.    
\end{itemize}

\subsubsection{HTTP Observers} \label{https observers.}
The extension employs three observers (listeners) running in the background (as long as the extension is running): 

\begin{enumerate}
\item \textbf{HTTP Before Send Request Observer.} 
This observer examines every HTTPS request that goes through the main address bar. The URL is either manually entered in the address bar by the user or automatically through URL redirection or through clicking on links. The examination occurs \textit{before} the request is sent, against the pre-defined domain names list. If the examined URL (e.g. mail.example.com/etc) belongs to any of the whitelisted domains (e.g. example.com), the extension enforces the TLS policy that is assigned to that domain. If the requested URL does not belong to any of the whitelsited domains, the extension enforces the browser\textquotesingle s default policy. After the policy enforcement, the request is sent. Note that the browser re-writes the configurations in real-time. Therefore, if the next URL does not belong to a whitelisted domain, the default policy will be re-enforced again.

\item \textbf{HTTP Response Header Observer.}
This observer examines every HTTP response header against a pre-defined header that we name it \quotes{\texttt{\justify strict-transport-security-config}}. A server that wishes to subscribe to our strict TLS policy must send this header in its HTTP response. Upon receiving, the browser interprets this as a request to add the domain to the whitelist in the \texttt{strict} configurations policy with a \texttt{blocking} error handling. Advertising security policies through the HTTP response header has been employed in the literature in other policies such as HSTS \cite{hstslist17}\cite{hsts17} to enforce HTTPS to protect against TLS stripping attacks, CSP \cite{csp10} to enforce trusted sources for page content scripts to protect against script injection attacks, and PKP header \cite{pinning15} to bind specific public keys to a website to protect against forged certificates. However, as stated earlier, the header advertisement method assumes a TOFU, i.e. the header is sent from an authentic server and not a man-in-the-middle. In addition, ideally such headers also contains a maximum-age parameter that specifies an age after which the header is expired, and the server needs to re-subscribe through the next header (in our mechanism header expiration implies removing the domain from the whitelist). For simplicity, in our proof-of-concept, our header consists of a name field only, without any fine-grained header parameters such as the maximum-age.

\item \textbf{HTTP Error Observer.} 
This observer is triggered when the request can not be processed due to lack of common TLS version or ciphersuite between the client and server. Our extension builds on the browser\textquotesingle s built-in error detection mechanism but we customise the browser error page if the error occurred for one of the whitelisted domains. Our extension detects the version or ciphersuites mismatch errors by observing the loaded documents (i.e. tabs) Uniform Resource Identifier (URI). We match every loaded tab\textquotesingle s URI against defined patterns that represent the Firefox\textquotesingle s version and ciphersuites mismatch errors URIs. In particular, we check if the loaded tab URI starts with \quotes{about:neterror} and contains either \quotes{SSL_ERROR_UNSUPPORTED_VERSION} or \quotes{SSL_ERROR_NO_CYPHER_OVERLAP}. Note that these patterns are vendor-specific. If a match is found, the extension extracts the URL that caused the error from the tab URI. Then, it examines the just extracted URL against the whitelist. If the URL belongs to a whitelisted domain name, the extension updates the tab with our extension\textquotesingle s customised error page according to the error handling mechanism that is assigned to the domain name. 
\end{enumerate}

\subsubsection{Error Message.} \label{fallback}
As described earlier, if the browser could not complete the handshake due to lack of common TLS version or ciphersuite with the server, a customised error page is shown to the user. Our mechanism employs two approaches for error handling: \texttt{blocking} and \texttt{active warning}. The browser selects the strategy based on the error handling mechanism that is assigned to the domain in the whitelist as described earlier in this section (see \quotes{Pre-Defined Error Handling Mechanism}). \par

In all cases, the error message is shown when the suspicious is higher than normal, based on the prior knowledge the browser has obtained either from the user or from the server about the sensitivity of the domain. It presents the user with a short message describing the reason of the error. If the error handling mechanism is \texttt{active warning} the message also contains two buttons. The first button is labeled \quotes{Restore Defaults}, and the second one is \quotes{Close}. The first button will relax the domain\textquotesingle s policy to the \texttt{default} policy and will try to connect again, through one click. This approach is similar to the Firefox built-in approach for handling version or ciphersuite mismatch error which presents \quotes{Restore Defaults} button that restores the Firefox\textquotesingle s default TLS versions and ciphersuites and reconnect, through one click. However, there is an intrinsic difference between our \texttt{active warning} error handling mechanism and the Firefox built-in mechanism. In our mechanism, the \quotes{Restore Defaults} button will change the configurations of the concerned domain only, and will not affect any other domain that the user may desire a \texttt{strict} TLS policy for. Thanks to our fine-grained TLS configurations concept that enables this feature. On the other hand, the \quotes{Restore Defaults} in the built-in Firefox warning will change the global configurations which will relax the configurations at a coarse-grained level and the new configurations will affect every connection. \par

Our warning message design is an initial prototype. Indeed, a further study with a focus on the usability aspect in addition to a user study needs its own space and we leave this for future work. 

\section{Other Methods for Policy Advertisement} \label{other methods}
There are other server-side policy advertisement methods that can be employed. In this section, we briefly describe some methods.
  
\begin{itemize}   
\item \textbf{Domain Name System (DNS) Record.} 
The DNS \cite{dns87} in conjunction with DNSSEC \cite{dnssec99} (the latter is to provide authentication) can include records for policy advertisement. This method has been proposed in HTTPSSR \cite{httpssr07}, a mechanism that advertises TLS support by a domain name through a DNS record, to protect against stripping attacks. If DNS in conjunction with DNSSEC is used for TLS configurations advertisement, it eliminates the TOFU issue and allows the configurations to be effective before the first connection request is sent since the DNS query is performed before the TLS request is sent. However, DNSSEC adoption might still be a barrier to rely on DNS for policy advertisement as noted in \cite{jackson08}.       
 
\item \textbf{Certificates.} The use of certificates to advertise policies has been proposed in PoliCert \cite{policert14}. PoliCert proposes a separate certificate for policies that has optional parameter that informs the browser about the server\textquotesingle s desired TLS minimum security level. The policy certificate method eliminates the TOFU issue since the certificate is signed by a trusted-third-party. However, it inherits the trusted-third-party complexity such as the cost since the domain owner needs to sign the policy by multiple CAs. Furthermore, similar to the HTTP header advertisement method, the policies can not be enforced before the first request is sent as the certificate needs to be received in a first connection.
\end{itemize}

\section{Limitations} \label{limitations}
In our proof-of-concept implementation there are few limitations: First, we used Add-on SDK (which is deprecated starting from Firefox 57.0) to perform the configurations re-writing which is not supported in Webextensions API. However, our present purpose is to demonstrate the feasibility of the concept. The configurations re-writing will not represent an issue if the mechanism got implemented at the browser source code level. Second, we do not consider measuring the performance at this stage. It can be measured if the mechanism is implemented at the browser source code level. As stated earlier, our scope in this paper is to propose and test the feasibility of the concept.

\section{Conclusion and Future Work} \label{conclusion}
Motivated by the experimental deployment of the coming version of TLS, TLS~1.3, we look at the problem of providing backward compatibility with legacy servers while maintaining security in web browsers. We propose a mechanism for fine-grained TLS security configurations in web browsers to augment browsers security and reduce the attack surface that exploits the client support for legacy versions, and non preferred or weak ciphersuites. Our proposal enables web browsers to learn about websites sensitivity and enforce optimal TLS configurations when connecting to sensitive websites while enforcing default configurations when connecting to the rest of the websites. This is an improvement over the \quotes{one-size-fits-all} coarse-grained TLS configurations mechanism that is used in all mainstream web browsers today. Our mechanism represents a middle-ground between optimal TLS configurations that might render many ordinary websites unreachable and default configurations that might be abused by attackers. We present our tool\textquotesingle s architecture and examine the feasibility of our proposal by implementing a proof-of-concept as a Firefox browser extension. We envision this mechanism as a built-in security feature in modern web browsers and as a standardised HTTP header that augment browsers security. Future work will focus on the usability aspect in addition to exploring new methods for server-based policy advertisement.

\section*{Acknowledgment}
The authors would like to thank Prof. Karthikeyan Bhargavan and Prof. Andrew Martin for useful feedback, Nicholas Moore for useful discussions on javascript, William Seymour and John Gallacher for proofreading. 

\FloatBarrier
\bibliographystyle{splncs03}
\bibliography{refs}
\end{document}